# CLUSTER GLOBALLY, REDUCE LOCALLY: SCALABLE EFFICIENT DICTIONARY COMPRESSION FOR MAGNETIC RESONANCE FINGERPRINTING

*Geoffroy Oudoumanessah*[1,2,3]  *Thomas Coudert*[1]  *Luc Meyer*[2]
*Aurelien Delphin*[1]  *Thomas Christen*[1]  *Michel Dojat*[1,2]  *Carole Lartizien*[3]  *Florence Forbes*[2]

[1] Univ. Grenoble Alpes, Inserm U1216, CHU Grenoble Alpes, Institut des Neurosciences, France
[2] Univ. Grenoble Alpes, Inria, CNRS, Grenoble INP, LJK, France
[3] Univ. Lyon, CNRS UMR 5220, Inserm U1294, INSA Lyon, UCBL, CREATIS, France

## ABSTRACT

With the rapid advancements in medical data acquisition and production, increasingly richer representations exist to characterize medical information. However, such large-scale data do not usually meet computing resource constraints or algorithmic complexity, and can only be processed after compression or reduction, at the potential loss of information. In this work, we consider specific Gaussian mixture models (HD-GMM), tailored to deal with high dimensional data and to limit information loss by providing component-specific lower dimensional representations. We also design an incremental algorithm to compute such representations for large data sets, overcoming hardware limitations of standard methods. Our procedure is illustrated in a magnetic resonance fingerprinting study, where it achieves a 97% dictionary compression for faster and more accurate map reconstructions.

## 1. INTRODUCTION

The development of medical imaging has led to data acquisition and production at much larger scales, for an increasing benefit to medical decision making. The exploitation of such large-scale data poses a number of challenges. A first challenge comes from the data processing scalability. Large-scale data may not be easily stored into memory or may be collected in a distributed manner from several sources, *e.g.* hospitals. Such a limited or distributed storage may limit the use of traditional methods, which load all the data into memory before running some optimization procedure. A second challenge comes from the data dimensionality. Across a wide range of medical applications, measured observations are high dimensional, *e.g.* magnetic resonance (MR) fingerprints [1], functional MR signals, neural network latent representations of images [2], *etc.* A typical difficulty is that the number of parameters for a model of such data can then easily exceed the number of observations, leading to estimation issues. In such high-dimensional settings, it is often possible to reduce the number of parameters by assuming that most of the information in the data can be captured and represented in a much lower dimensional subspace. Classical techniques include principal component analysis (PCA), probabilistic principal component analysis (PPCA) [3], factor analysers (FA), and newer methods such as diffusion maps [4]. More flexible approaches are also based on mixtures of the previous ones, such as mixtures of factor analysers (MFA) [5] and mixtures of PPCA (MPPCA) [6]. Another mixture approach is called HD-GMM in [7] for High Dimensional Gaussian Mixture Models. It encompasses many forms of MFA and MPPCA and generalises them, see also [7] for a review on high dimensional clustering via mixtures. However, most of these methods are designed for batch data and are thus sensitive to hardware limits such as memory, which restricts the amount of data they can process or the type of medical devices on which they can be usefully embedded. As a simple solution, most implementations downsample data sets before processing, potentially loosing useful information. Another approach is to design incremental, also referred to as online, variants handling data sequentially in smaller groups. A number of incremental approaches exist for dimension reduction techniques, see the recent SHASTA-PCA [8] and [9] for a review. To our knowledge, much fewer solutions exist for mixtures. Estimation of such models is generally based on maximum likelihood estimation via the Expectation-Maximization (EM) algorithm [10]. We can mention a preliminary attempt for an incremental MPPCA based on heuristic approximations of the EM steps [11]. In this work, considering data both large in size and dimension, our proposal is twofold. Building on HD-GMM, originally developed for high dimensional clustering and density estimation, we show how they can be used to compress high dimensional data into several reduced size data subsets. We then derive a new incremental algorithm, based on a principled EM framework, to learn such a model from very large data sets. We demonstrate the effectiveness of our approach on a MR fingerprinting (MRF) study, allowing to go far beyond the simulations resolution and size used in current implementations and to reconstruct a larger

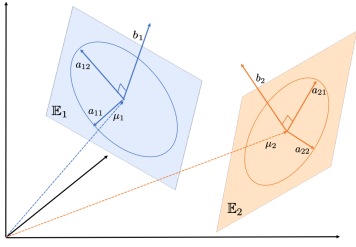

**Fig. 1**: HD-GMM schematic illustration, $M=3, d=2, K=2$.

| | SNR (dB) | $d=8$ | $d=10$ | $d=15$ |
|---|---|---|---|---|
| | - | 0.35 | 0.33 | 0.30 |
| SVD [12] | 15 | 0.40 | 0.35 | 0.32 |
| | 5 | 0.44 | 0.43 | 0.40 |
| | - | 0.031 | 0.030 | **0.028** |
| HD-GMM | 15 | 0.032 | 0.031 | **0.028** |
| | 5 | 0.16 | 0.15 | **0.10** |
| Size (Go) | | **0.30** | 0.38 | 0.57 |

**Table 1**: Compressed dictionary MAEs and sizes. The lower the better, best values in bold.

number of MR parameter maps with an improved accuracy.

## 2. DIVIDE & CONQUER REDUCTION OF LARGE DATA VOLUMES

**Identifying group-wise subspaces.** HD-GMM assume that the observations are *i.i.d.* realizations of a random variable $\mathbf{y}$ which follows a Gaussian mixture model with $K$ components,

$$p(\mathbf{y}; \boldsymbol{\theta}) = \sum_{k=1}^{K} \pi_k \mathcal{N}_M(\mathbf{y}; \boldsymbol{\mu}_k, \boldsymbol{\Sigma}_k), \quad (1)$$

where $\mathcal{N}_M$ denotes the $M$-dimensional Gaussian distribution and $\boldsymbol{\mu}_k, \boldsymbol{\Sigma}_k, \pi_k$ are respectively the $k$th component mean, covariance matrix and weight. An efficient parameters reduction can be obtained by using the eigendecomposition $\boldsymbol{\Sigma}_k = \boldsymbol{D}_k \boldsymbol{A}_k \boldsymbol{D}_k^T$, where $\boldsymbol{D}_k$ contains the eigenvectors $\{\boldsymbol{d}_1, \ldots, \boldsymbol{d}_M\}$ of $\boldsymbol{\Sigma}_k$, and $\boldsymbol{A}_k$ is a diagonal matrix with its eigenvalues. In HD-GMM, each $\boldsymbol{A}_k$ consists of only $d_k + 1$ different eigenvalues $\boldsymbol{A}_k = \text{diag}(a_{k1}, \ldots, a_{kd_k}, b_k, \ldots, b_k)$, with $a_{kj} > b_k$, for $j = 1:d_k$, and where $d_k \in \{1, \ldots, M-1\}$ is *a priori* unknown but fixed in our work to a user decided dimension $d$. When $b_k$ is negligible, this parameterization means that a group-specific subspace $\mathbb{E}_k$, parameterised by the $d$ eigenvectors associated to the first $d$ eigenvalues $\{a_{k1}, \ldots, a_{kd}\}$, captures the main cluster shape (see Figure 1 for an illustration). The model parameters are then $\boldsymbol{\theta} = \{\boldsymbol{\theta}_k, k = 1:K\}$ with $\boldsymbol{\theta}_k = \{\pi_k, \boldsymbol{\mu}_k, \boldsymbol{A}_k, \boldsymbol{D}_k\}$. They can be estimated using an EM algorithm. When $d \ll M$, a significant computation gain can be achieved. Let $\widetilde{\boldsymbol{D}}_k$ consist of the $d$ first columns of $\boldsymbol{D}_k$ supplemented by $(M-d)$ zero columns and $\overline{\boldsymbol{D}}_k = (\boldsymbol{D}_k - \widetilde{\boldsymbol{D}}_k)$. Then, $P_k(\mathbf{y}) = \widetilde{\boldsymbol{D}}_k \widetilde{\boldsymbol{D}}_k^T (\mathbf{y} - \boldsymbol{\mu}_k) + \boldsymbol{\mu}_k$ and $P_k^\perp(\mathbf{y}) = \overline{\boldsymbol{D}}_k \overline{\boldsymbol{D}}_k^T (\mathbf{y} - \boldsymbol{\mu}_k) + \boldsymbol{\mu}_k$ are the projections of $\mathbf{y}$ on $\mathbb{E}_k$ and its orthogonal space $\mathbb{E}_k^\perp$. The main EM computations involve quadratic quantities $(\mathbf{y} - \boldsymbol{\mu}_k) \boldsymbol{\Sigma}_k^{-1} (\mathbf{y} - \boldsymbol{\mu}_k)$ which can be equivalently written as,

$$\|\boldsymbol{\mu}_k - P_k(\mathbf{y})\|^2_{\widetilde{\boldsymbol{\Sigma}}_k^{-1}} + \frac{1}{b_k} \|\mathbf{y} - P_k(\mathbf{y})\|^2, \quad (2)$$

where $\|.\|^2_{\widetilde{\boldsymbol{\Sigma}}_k^{-1}}$ is the norm defined by $\|\mathbf{y}\|^2_{\widetilde{\boldsymbol{\Sigma}}_k^{-1}} = \mathbf{y}^T \widetilde{\boldsymbol{\Sigma}}_k^{-1} \mathbf{y}$ with $\widetilde{\boldsymbol{\Sigma}}_k^{-1} = \widetilde{\boldsymbol{D}}_k \boldsymbol{A}_k^{-1} \widetilde{\boldsymbol{D}}_k^T$. Expression (2) uses the definitions of $P_k, P_k^\perp$ and $\|\boldsymbol{\mu}_k - P_k^\perp(\mathbf{y})\|^2 = \|\mathbf{y} - P_k(\mathbf{y})\|^2$. The gain comes from the fact that (2) does not depend on $P_k^\perp$ and thus does not require the computation of the $(M - d)$ latest columns of $\boldsymbol{D}_k$, the eigenvectors associated to the smallest eigenvalues. Similarly, determinants can be efficiently computed as $\log(|\boldsymbol{\Sigma}_k|) = (\sum_{j=1}^{d} \log(a_{kj})) + (M - d)\log(b_k)$. This parameterization allows to handle high dimensional data in a computationally efficient way. However, it does not provide an actual lower dimensional representation of the data. While such a reduced-dimensional representation may often not be needed, it may be crucial to deal with hardware or software limitations. Originally, HD-GMM have not been designed for this situation, but we describe next how they can be further exploited as a dimension reduction technique.

**Cluster-wise dimension reduction.** As clustering models, for any possible observation $\mathbf{y}$, HD-GMM provide a probability $r_k(\mathbf{y})$ that $\mathbf{y}$ is assigned to cluster $k$ for each $k = 1:K$. Denote $\widetilde{\boldsymbol{D}}_k^*$ the $M \times d$ matrix built with the $d$ first columns of $\boldsymbol{D}_k$. A reduced-dimensionality representation $\widehat{\mathbf{y}}_k$ of $\mathbf{y}$ can be obtained, for each of the $K$ different subspaces, by computing the scalar products of a centered $\mathbf{y}$ with the columns of $\widetilde{\boldsymbol{D}}_k^*$. It comes $\widehat{\mathbf{y}}_k = S_k(\mathbf{y}) = \widetilde{\boldsymbol{D}}_k^{*T}(\mathbf{y} - \boldsymbol{\mu}_k)$, while its reconstruction $\widetilde{\mathbf{y}}_k$ in the original space is given by $\widetilde{\mathbf{y}}_k = \widetilde{\boldsymbol{D}}_k^* \widehat{\mathbf{y}}_k + \boldsymbol{\mu}_k$. In practice, it is reasonable to use as a reduced-dimensionality representation of $\mathbf{y}$ only the one corresponding to the most probable group $k$, *i.e.* with the highest $r_k(\mathbf{y})$. In this setting, HD-GMM acts as a divide-and-conquer paradigm by clustering the data into $K$ clusters and allowing cluster-specific data reduction. The divide step allows a much more effective reduction than if a single subspace is considered for the whole data set. In the conquer step, little information is lost, as for any new observation $\mathbf{y}$, cluster assignment probabilities $r_k(\mathbf{y})$ can be straightforwardly computed to decide on the best reduced representation to be used. However, for subsequent processing, it is important to keep track of clustering information for each observation. The reduced representations cannot be pooled back altogether, as they would be likely to become impossible to distinguish across clusters.

**Incremental learning for large data volumes.** In practice, most approaches lie on optimization procedures requiring all data to be loaded in a single batch. Batch sizes are then limited by resource constraints, so that very large data sets need either to be downsampled or to be handled in an incremental manner, *i.e.* with smaller data subsets processed sequentially. Incremental versions of EM exist and can be adapted to our setting. As an archetype of such algorithms, we consider the online EM of [13] which belongs to the family of stochastic approximation algorithms. We refer to [13] for details on the main assumptions required and the online EM iteration. When applied to mixture models, it can be shown [14], that it is enough to check that each mixture component has an exponential family form. For the HD-GMM parameterization, omitting the cluster index $k$, the exponential form of $\mathcal{N}_M(\mathbf{y}; \boldsymbol{\mu}, \boldsymbol{\Sigma})$ is given by $h(\mathbf{y}) \exp\left(s(\mathbf{y})^T \boldsymbol{\phi}(\boldsymbol{\mu}, \boldsymbol{\Sigma}) - \psi(\boldsymbol{\mu}, \boldsymbol{\Sigma})\right)$ where $s$ and $\boldsymbol{\phi}$ are vectors with respective elements ($vec(\cdot)$ is the vectorization operator of a matrix) $\left[\mathbf{y}, \text{vec}(\mathbf{y}\mathbf{y}^T), \mathbf{y}^T \mathbf{y}\right]$ and

$$\left[\sum_{j=1}^{d}(\tfrac{1}{a_j} - \tfrac{1}{b})\boldsymbol{d}_j\boldsymbol{d}_j^T\boldsymbol{\mu} + \tfrac{1}{b}\boldsymbol{\mu}, \tfrac{1}{2}\sum_{j=1}^{d}(\tfrac{1}{b} - \tfrac{1}{a_j})\text{vec}(\boldsymbol{d}_j\boldsymbol{d}_j^T), -\tfrac{1}{2b}\right]$$

and $\psi(\boldsymbol{\mu}, \boldsymbol{\Sigma})$ is equal to

$$\tfrac{1}{2}\sum_{j=1}^{d}\left[(\tfrac{1}{a_j} - \tfrac{1}{b})\boldsymbol{\mu}^T\boldsymbol{d}_j\boldsymbol{d}_j^T\boldsymbol{\mu} + \log(a_j)\right] + \tfrac{1}{2b}\boldsymbol{\mu}^T\boldsymbol{\mu} + \tfrac{(M-d)}{2}\log(b).$$

The form of $h(\cdot)$ is irrelevant for the computations. It follows an online EM algorithm which is closed-form except for the update of $\widehat{\boldsymbol{D}}^*$, which is estimated using a Riemanian optimization framework in the setting where $M >> d$ [15].

## 3. LARGE SCALE MR FINGERPRINTING (MRF)

MRF [1] allows the simultaneous acquisition and reconstruction of multiple tissue properties maps, see [16, 17] for recent reviews. In the original *matching* approach, maps reconstruction is based on the search for the best match between an observed signal and a dictionary of simulated signals (fingerprints). As an alternative, learning approaches have been studied via various neural network (NN) architectures, but improvement has been demonstrated mainly for standard $T_1$ and $T_2$ relaxometry parameters estimation and only up to 3 parameters in the dictionary, see *e.g.* [18] or Table 2 in [17]. To now, none of these approaches has shown real scalability properties with respect to the number of parameters to be reconstructed [18]. The main issue is that the size of the simulated dictionary increases exponentially with the number of parameters to be reconstructed. In this context, [19] showed that a dictionary-based Bayesian learning approach was more accurate and less demanding, in terms of dictionary size, than conventional dictionary matching and some NN solutions. We build on this work by using online HD-GMM estimation to handle an extensive high-resolution dictionary. This offers a gain in data storage, when traditional dictionary matching is used, but also in accuracy and speed of map reconstructions, when Bayesian or NN methods are considered. In a matching approach, our proposal is similar in spirit to [20], which distribute the matching cost into smaller matching tasks. However [20] does not adapt straightforwardly to learning-based or NN approaches. For this illustration, we use the method introduced in [21], to simulate a dictionary for various ranges of relaxometry parameters $T_1$, $T_2$, $T_2^*$, and magnetic field parameters $\delta f$ and $B_1$. In particular, transverse relaxation time $T_2^*$ is a MR tissue property that provides insight into underlying tissue physiology and pathology, making this MR parameter a widely used biomarker of several clinical diseases. The dictionary is made of 4.750M signals in dimension 260, for a storage cost of more than 20Gb. Gaussian noise is often added to signals using different SNR values [18]. Table 1 first shows the efficiency of different dictionary compression strategies. Mean absolute error (MAE) values are computed, for different noise levels, between the original signals and their denoised reconstructions after compression. In all cases, HD-GMM show much lower compression losses than the reference SVD method [12], achieving better reconstructions even with a dimension reduced to $d = 8$, about half smaller than SVD ($d = 15$). For HD-GMM their are two hyperparamters namely the number of components $K$ and the reduced dimension $d$. Both can be chosen using a Bayesian information criterion (BIC), which in our case gives $K = 30$ and $d = 10$. For initializing our EM algorithm, we use the heuristic mentioned in [22] Section 3.5, which provides good performance for our approach too. Parameters maps reconstruction is then illustrated on $T_1$, $T_2$, $T_2^*$ in Figure 2. We first perform a standard matching, referred to as *full matching*, based on inner products between *in vivo* acquired MR signals and simulated signals from the original high dimensional dictionary (Figure 2 line 1). The *in vivo* acquisition was performed on one healthy volunteer with a Philips 3T Achieva dStream MRI at the IRMaGe facility. As ground truth maps are not available, full matching maps often serve as reference maps, although this has obvious limits in terms of hardware and robustness to noise. Matching results are then shown for reduced dictionary representations, using either SVD with $d = 10$ (Figure 2 line 2) or HD-GMM with $d = 10$ and $K = 30$. As learning-based alternatives, maps obtained with DRONE [23], a 4 layers fully connected NN, and with a combination of the HD-GMM reduction and the Bayesian learning model referred to as GLLiM [19], are also shown (Figure 2 line 3 and 5 respectively). The GLLiM model corresponds to a regression model for which a number of Gaussian components needs to be chosen and is set to $K_g = 50$. Figure 2 clearly shows that both SVD and DRONE provide unsatisfying maps, *e.g.* for $T_2$ and $T_2^*$. The HD-GMM approach is visually similar to full matching. Its combination with GLLiM provides close to full matching and more spatially homogeneous maps, despite some remaining *shim* artifacts that we interpret as learning bias due to the high cross-correlation between $\delta f$ nominal value and the resulting $T_2^*$ value. HD-GMM superior performance is then confirmed quantitatively in Table 3 where the MAE, over all voxels, with respect to full matching is shown for all 5 parameters. In Table 3, MAEs are computed with respect to imperfect still noisy full matching maps, which explains why the GLLiM variant of HD-GMM does not provide the lowest MAE despite more satisfying maps in terms of spatial homogeneity. Another quantitative comparison is provided in Table 2, with $T_1$, $T_2$ and $T_2^*$ mean values over voxels, respectively in white and grey matter ROIs, delineated on $T_1$ maps. When compared to ranges in healthy subjects as provided in [21], HD-GMM variants show more in range values than most other methods. In Python with Jax library (link to our code upon acceptation), full matching takes 11s, SVD matching takes 4s and the HD-GMM variants take less than 0.4s on Nvidia V100 GPU. However, in more realistic situations, calculations are performed locally by a medical practitioner on a CPU. On an Apple M2 Pro CPU, full matching takes 1 to 2 hours, while HD-GMM variants take only 2min.

| Parameter | ROI | Full matching | SVD | DRONE | HD-GMM | HD-GMM+GLLiM | Literature [21] |
|---|---|---|---|---|---|---|---|
| $T_1$(ms) | WM | $868 \pm 2$ | $905 \pm 2$ | $850 \pm 2$ | $847 \pm 2$ | $834 \pm 1$ | 690-1100 |
|  | GM | $1373 \pm 7$ | $1400 \pm 7$ | $1272 \pm 6$ | $1360 \pm 7$ | $1337 \pm 7$ | 1286-1393 |
| $T_2$(ms) | WM | $49 \pm .1$ | $87 \pm .2$ | $50 \pm .2$ | $55 \pm .2$ | $55 \pm .2$ | 56-80 |
|  | GM | $73 \pm 1$ | $118 \pm 1$ | $77 \pm 1$ | $81 \pm 1$ | $81 \pm .1$ | 78-117 |
| $T_2^*$(ms) | WM | $46 \pm .1$ | $23 \pm .5$ | $29 \pm 24$ | $51 \pm .2$ | $51 \pm .2$ | 45-48 |
|  | GM | $46 \pm .4$ | $30 \pm .4$ | $27 \pm 33$ | $55 \pm .6$ | $51 \pm .1$ | 42-52 |

**Table 2**: Mean $T_1, T_2, T_2^*$ values with 99% confidence in white (WM) and grey (GM) matters ROIs. Out of range values in orange and red, red is further out.

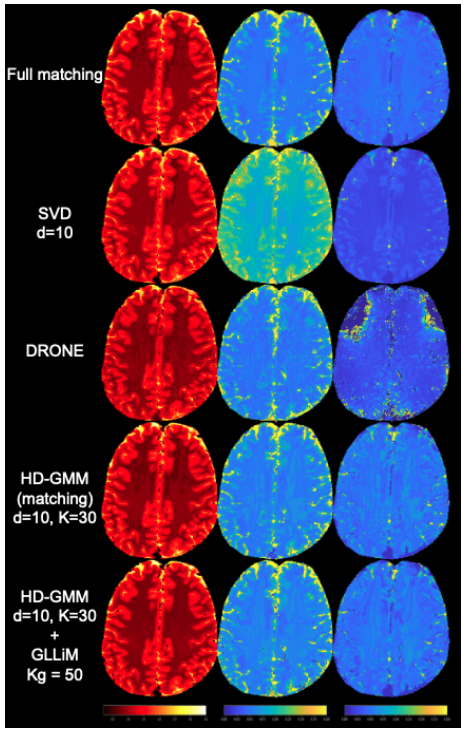

**Fig. 2**: $T_1, T_2, T_2^*$ maps (columns) for various methods (lines).

| Method | $T_1$ (s) | $T_2$ (s) | $T_2^*$ (s) | $\delta f$ (Hz) | $B_1$ sensitivity |
|---|---|---|---|---|---|
| DRONE [23] | 0.14 | 0.022 | 0.096 | 5.0 | 5.5 |
| SVD [12] | **0.056** | 0.047 | 0.018 | 1.4 | 0.1 |
| HD-GMM + matching | 0.058 | **0.016** | **0.010** | **1.0** | **0.04** |
| HD-GMM + GLLiM [19] | 0.081 | 0.019 | 0.012 | 1.3 | 0.05 |

**Table 3**: MAEs over voxels with respect to full matching, for DRONE, SVD ($d = 10$), and HD-GMM ($d = 10$, $K = 30$, $K_g = 50$). Best values in bold.

## 4. CONCLUSION AND FUTURE WORK

By equipping high-dimensional Gaussian mixtures models (HD-GMM) with a dimension reduction procedure and incremental estimation of their parameters, we showed that HD-GMM could scale to both very large and high-dimensional data sets. These models can act as a divide-and-conquer paradigm by initially clustering large data volumes and then performing cluster-specific dimension reduction. The clustering structure allows to achieve larger dimension reduction for the same information loss. This ability was showcased on an MRF study using more parameters and more dictionary entries than in standard MRF settings. HD-GMM can use more informative simulations for more accurate parameters maps and thus provide a promising direction towards unleashing the full power of MRF. Future work includes testing further HD-GMM on extensive dictionaries for an increased number of parameters. However for ultra-high MRF dictionaries, the sizes of the obtained sub-dictionaries with HD-GMM may be too high to be efficiently handled via traditional matching. Coupling HD-GMM and GLLiM regression model of [19] would then provide a more tractable solution. More generally, other medical imaging applications can benefit from the scalability and flexibility on the proposed pipeline. HD-GMM allow an efficient use of higher dimensional information as can be extracted from NN latent representations, e.g. [2].